\documentclass[showpacs,showkeys,12pt,preprint,preprintnumbers,nofootinbib,
groupedaddress,superscriptaddress,amsmath,amssymb,axodraw]{revtex4}

\usepackage[dvips]{color}
\usepackage{graphicx}
\usepackage{epsfig}    
\usepackage{dcolumn}
\usepackage{bm}
\usepackage{here}

\def\Journal#1#2#3#4{{#1} {#2} (#4) #3}

\def\NPB{{Nucl. Phys.} B}
\def\PLB{{Phys. Lett.}  B}
\def\PRL{Phys. Rev. Lett.}
 
\def\PRD{{Phys. Rev.} D}
\def\ZPC{{Z. Phys.} C}
\def\EPC{{Eur. Phys. J.} C}

\def\JHEP{J.~High Energy Phys.}
\def\MPLA{{Mod.~Phys.~Lett.} A}

\allowdisplaybreaks[4]

\newcommand{\bspace}{\!\!\!\!}
\newcommand{\met}{\mbox{$E{\bspace}/_{T}$}}

\newcommand{\gsim}{\mbox{ \raisebox{-1.0ex}{$\stackrel{\textstyle >}
{\textstyle \sim}$ }}}
\newcommand{\lsim}{\mbox{ \raisebox{-1.0ex}{$\stackrel{\textstyle <}
{\textstyle \sim}$ }}}

\begin{document}

\title{Potential for measuring the $H^\pm W^\mp Z^0$ vertex 
from $WZ$ fusion \\at the Large Hadron Collider }

\author{Eri Asakawa}
\email{eri@post.kek.jp}
\affiliation{Theory Group, KEK, \\ Tsukuba, Ibaraki 305-0801, Japan}

\author{Shinya Kanemura}
\email{kanemu@sci.u-toyama.ac.jp}
\affiliation{Department of Physics, University of Toyama, 
 3190 Gofuku, Toyama 930-8555, Japan}

\author{Junichi Kanzaki}
\email{Junichi.Kanzaki@cern.ch}
\affiliation{Institute of Particle and Nuclear Studies,
  KEK, \\ Tsukuba, Ibaraki 305-0801, Japan}

\preprint{KEK-TH-1122}
\preprint{UT-HET 005}



\begin{abstract}

 We investigate the possibility of measuring the $H^\pm W^\mp Z^0$
 vertex from the single $H^\pm$ production process via $WZ$ fusion
 at the CERN Large Hadron Collider (LHC).
 This vertex strongly depends on the structure of the Higgs sector
 in various new physics scenarios, so that its measurement can be useful to
 distinguish the models. 
 A signal and background simulation under the expected detector
 performance at the LHC is done for the processes of
 $pp \to W^\pm Z^0 X \to H^\pm X \to tbX$ and
 $pp \to W^\pm Z^0 X \to H^\pm X \to W^\pm Z^0 X$,
 and the required magnitudes of the $H^\pm W^\mp Z^0$ vertex
 for observation are evaluated.
 It is found that although the loop induced $H^\pm W^\pm Z^0$ vertex in  
 multi-Higgs doublet models cannot be measurable, the latter process
 can be useful to test the model with a real and a
 complex triplets. 
  
\end{abstract}

\maketitle

\section{Introduction}

The idea of spontaneous breaking of the electroweak
gauge symmetry will soon be tested directly at the CERN Large
Hadron Collider (LHC)~\cite{atlas-cms}. 
There, it is expected that a Higgs boson of the
standard model (SM) can be detected in a wide range of the mass,
which is the last unknown parameter of the model. 
On the other hand, it is possible that the
structure of the Higgs sector takes a non-minimal form
with additional isospin-singlets, doublets, triplets etc. 
In fact, there is a variety of new physics scenarios 
which deduce extended Higgs sectors in the low energy
effective theory.

In extended Higgs models, 
additional scalar bosons
like charged and CP-odd Higgs bosons are predicted.  
There are plenty of studies on production mechanisms of such extra Higgs bosons.
At the LHC, charged Higgs bosons $H^\pm$ in two Higgs doublet models (2HDMs),
including the minimal supersymmetric standard model (MSSM), are produced via 
$gg \rightarrow H^\pm tb$ (or $gb \rightarrow H^\pm t$)~\cite{gb},
$gg(q\overline{q}) \rightarrow H^+ H^-$~\cite{pairpro} and
$gg(q\overline{q}) \rightarrow H^\pm W^\mp$~\cite{HW} 
when $m_{H^\pm} \gsim m_t$, while they are produced via
$t \rightarrow b H^+$ for $m_{H^\pm} \lsim m_t-m_b$~\cite{atlas-cms, tdecay}.
These processes are primarily important for discovery of charged Higgs bosons
in such models. On the other hand,
if the Higgs sector includes greater multiplets
than doulbets, the production and the decay of the singly-charged Higgs bosons can be 
completely different from those in the 2HDMs.
Therefore, it is valuable to study more observables which can be measured
at future experiments in order to
determine the Higgs sector.

For the structure of the Higgs sector,
the current electroweak data provide important hints.  
In particular, all the Higgs models must
respect the experimental fact that the electroweak rho parameter
($\rho$) is very close to unity.
It is well known that the tree-level prediction of $\rho=1$ is a
common feature of Higgs models with only doublets (and singlets)~\cite{HWZ}.
On the contrary, the rho parameter data give a strong constraint
on the models which additionally include triplets or greater
representations of the isospin $SU(2)$ gauge symmetry, because
in these cases the predicted rho parameter is generally not unity
already at the tree level~\cite{hhg}.
There are two simple possibilities to satisfy this constraint
in such models.
First, these multiplets but doublets
do not contribute to the electroweak spontaneous
symmetry breaking much, and their vacuum expectation values are
sufficiently small as compared to those of doublets~\footnote{
This type of models includes
the Left-Right symmetric model~\cite{LR}, 
the Littlest Higgs model~\cite{LittlestHiggs,LH-ph} 
and some extra dimension models~\cite{extraD-triplet}
which predict an additional complex triplet. }.
Second, we can consider models arranging the custodial $SU(2)$
symmetry among these additional multiplets, so that the models
then naturally impose $\rho=1$ at the tree level~\cite{TRIP,cg,thetaH,logan}.

After the discovery of extra Higgs bosons, the Higgs sector
must be explored by measuring their masses and various coupling constants etc.
For the exploration of the global symmetry structure of the Higgs sector, 
the coupling of singly-charged Higgs bosons with the weak gauge 
bosons, $H^\pm W^\mp Z^0$, is of particular importance~
\cite{HWZ,HWZ-2HDM0,HWZ-2HDM1,HWZ-2HDM2,Arhrib,AKprev}.
It can appear at the tree level in models with scalar triplets, 
while it is induced at the loop level in multi scalar doublet models.
In Ref.~\cite{AKprev}, the new physics predictions to the $H^\pm W^\mp
Z^0$ are summarized and possible size of the signal cross section in each
model is evaluated.
The vertex is expressed by 
\begin{align}
i \Gamma_{H^\pm W^\mp Z}(p_W, p_Z^{}; \lambda_W^{}, \lambda_Z^{}) =
 i g m_W V_{\mu\nu} 
\epsilon_W^{\ast\mu}(p_W, \lambda_W) 
\epsilon_Z^{\ast\nu}(p_Z, \lambda_Z),
\end{align}
where $g$ is the weak gauge coupling, $m_W^{}$ is the mass of 
the weak boson $W^\pm$, and $\epsilon_V^{\ast\mu}(p_V, \lambda_V)$ 
($V=W$ and $Z$) are polarization vectors for 
the outgoing weak gauge bosons with the momentum $p_V^{}$ and 
the helicity $\lambda_V^{}$. Here, $V_{\mu\nu}^{}$ is 
decomposed in terms of three form factors 
as
\begin{align}
V_{\mu\nu} = F g_{\mu\nu}
            +\frac{G}{m_W^2}{p_Z^{}}_\mu {p_W^{}}_\nu
            +\frac{H}{m_W^2}\epsilon_{\mu\nu\rho\sigma}
             {p_Z^{}}^\rho {p_W^{}}^\sigma, \label{eq:HWZ}
\end{align}
where the antisymmetric tensor $\epsilon_{\mu\nu\rho\sigma}$ 
is defined so as to satisfy $\epsilon_{0123}^{}=-1$.
The values of $F$, $G$ and $H$ depend on 
the detail of the model.   The leading
contribution is well described by $F$, and its magnitude is
directly related to the structure of the extended Higgs sector
under global symmetries\cite{hhg,HWZ}.  

In this paper, we consider testing the 
$H^\pm W^\mp Z^0$ coupling via the single charged Higgs
boson production in the $WZ$ fusion process at the LHC.
In general, neutral Higgs production
processes via vector boson fusion (VBF) have an advantage over
the other production processes, 
because the signal can be completely  reconstructed
and produced jets are suppressed 
in the central region due to a lack of color flow 
between the initial state quarks~
\cite{atlas-cms,VBFheavy,VBFmiddle,zeppenfeld-ww,vbf}.
The same benefit can also be applied to $H^\pm$ production
from $WZ$ fusion.
A signal and background simulation under the expected detector
performance at the LHC is performed for the processes of
$pp \to W^\pm Z^0 X \to H^\pm X \to tbX$ and
$pp \to W^\pm Z^0 X \to H^\pm X \to W^\pm Z^0 X$.
We estimate the minimum value of the form factor $|F|^2$ above which
the signal significance can be substantial for a possibility
of detection at the LHC. We then
discuss which new physics models can be tested via these
processes.

\section{Possible values of the $H^\pm W^\mp Z^0$ vertex in several models}

We here shortly summarize typical values of the form factor $F$
in several new physics models\cite{AKprev}; i.e.,
2HDMs (including the MSSM)\cite{HWZ-2HDM0,
HWZ-2HDM1,HWZ-2HDM2,Arhrib,eeHW-2HDM1,eeHW-2HDM2,singleH+,eeHW-MSSM},
the Littlest Higgs model\cite{LittlestHiggs,LH-ph}, and the model
with a real and a complex triplet fields\cite{TRIP,cg,thetaH}.

In the 2HDM\footnote{We consider Model II 2HDM\cite{hhg}.},
there are two sources to enhance the 
loop-induced form factors; i.e., the contribution from 
the top-bottom loop and those from the Higgs-boson loop.
The values of $F^{(t-b \;\rm loop)}$ are given by 
$F^{(t-b \;\rm loop)} \simeq 0.01 \cot\beta$; i.e.,
$|F^{(t-b \;\rm loop)}|^2        
\simeq 10^{-3}, 10^{-4}$ and
$10^{-5}$ for $\tan\beta=0.3,1$ and $3$, respectively,
where $\tan\beta$ is the ratio of vacuum expectation values
for neutral components of the two Higgs doublets.
The contribution of the Higgs boson loop can be 
important for $\tan\beta \gtrsim 3$, where the top-bottom 
loop contribution becomes suppressed because of the smaller 
Yukawa couplings\cite{eeHW-2HDM1}. 
The Higgs loop effect on
 $F$ is constrained from perturbative unitarity\cite{pu,2hdm_pu} as 
$|F^{({\rm bosonic\; loop})}|^2 \lesssim 10^{-5}$ 
for $3 \lesssim \tan\beta \lesssim 10$. 
Therefore, 
the value of $|F^{\rm (2HDM)}|^2$ can be at most 
$ \sim 10^{-3}$, $10^{-4}$ and  
$10^{-5}$ for $\tan\beta=0.3$, $1$ and $3-10$, respectively. 
In the MSSM, the loop effect of super partner 
particles can enhance the vertex especially 
in the moderate values of $\tan\beta$.
However, its magnitude is not large and
at most less than $10^{-5}$\cite{eeHW-MSSM}  for $\tan\beta \gtrsim 3$.

In models with triplets, the $H^\pm W^\mp Z^0$ vertex 
generally appears at the tree level. 
A common feature of the tree level contribution 
to the form factor $F$ is the fact that 
it is proportional to the vacuum expectation values (VEVs) of the triplet field\cite{hhg,TRIP},
$F   \propto  v'/v$, 
where $v$ and $v'$ represent the VEVs of the doublet 
and the triplet in the model, respectively. 
When more than one triplet appear in the model, 
$v'$ should be taken as the combination of the VEVs for them. 
In the Littlest Higgs model, an additional complex triplet field appears in the 
effective theory. The electroweak data indicate 
$1 \lesssim v' \lesssim 4$ GeV for $f=2$ TeV\cite{dawson}. 
We here consider the case with $m_h=115$ GeV,  $f=1$ TeV (2 TeV), 
$v' = 5$ GeV ($4$ GeV), and $m_{H^\pm}^{}=700$ GeV (1.56 TeV)
as a reference: i.e., 
the value of the form factor $F$ is 
$|F^{\rm (LLH)}|^2 \simeq 0.0085$ (0.0054).

In the model with additional real and complex triplet fields, 
the rho parameter can be set to be unity at the tree level, 
by imposing the custodial symmetry; i.e.,
$v_{\rm r}' = v_{\rm c}'(=v')$, 
where $v_{\rm r}'$ and $v_{\rm c}'$ are respectively 
the VEVs of the real and the complex triplet field\cite{TRIP,cg,thetaH}.
The strongest experimental bound on $v'/v$ 
comes from the $Z b \bar b$ result.  
The limits at 95\% CL are $\tan\theta_H^{} \lesssim 0.5, 1$ and $1.7$
for the mass of the three-plet field of $SU(2)_V$ which do not couple to
$W^\pm$ and $Z^0$
to be $0.1,0.5$ and $1$ TeV, respectively\cite{logan},
where $\sin\theta_H^{}=\sqrt{8  v'^2/(v^2+8 v'^2)}$.
These limits imply $|F^{\rm (triplet)}|^2 = 0.26-0.97$.

\section{Signal and backgrounds}
\label{Sec:predictions}

\subsection{Charged Higgs production via vector boson fusion}

VBF is a pure electroweak process without color flow in the central
region. The process naturally leads to high transverse momentum ($P_T$)
tag jets in the forward region and allow to observe small hadronic
activity in the central region except for jets from produced Higgs boson.
Hence it is possible to observe small electroweak signal rates
in a region of phase space not very populated by QCD background events.
Such an advantage of VBF has been used for neutral
Higgs boson production for intermediate and greater masses of Higgs
bosons~\cite{atlas-cms,VBFheavy,VBFmiddle,zeppenfeld-ww,vbf}. 
This can also be applied for charged
Higgs boson production from $WZ$ fusion, which we discuss in this paper.

In Fig. 1, the hadronic cross section of 
$pp \to W^{\pm\ast} Z^{0\ast} X \to H^\pm X$ at the LHC 
($\sqrt{s}=14$ TeV) 
is shown as a function of mass $m_{H^\pm}^{}$ of the charged Higgs boson~\cite{AKprev}. 
The form factors $F$, $G$ and $H$ of the $H^\pm W^\mp Z^0$ 
vertex are set to be 1, 0 and 0, respectively. 
The magnitude of $\sigma^{|F|^2=1}$ can be about 
$3900$ fb for $m_{H^\pm}^{}=200$ GeV, $470$ fb for $m_{H^\pm}^{}=500$ GeV, 
$210$ fb for $m_{H^\pm}^{}=700$ GeV and 
$150$ fb for $m_{H^\pm}^{}=800$ GeV. 
The prediction of the cross section in each new physics scenario can be obtained
by rescaling the value of $F$. The typical values of $F$
in several models are summarized in Sec.~II.

\begin{figure}[hth]
\includegraphics[width=10cm]{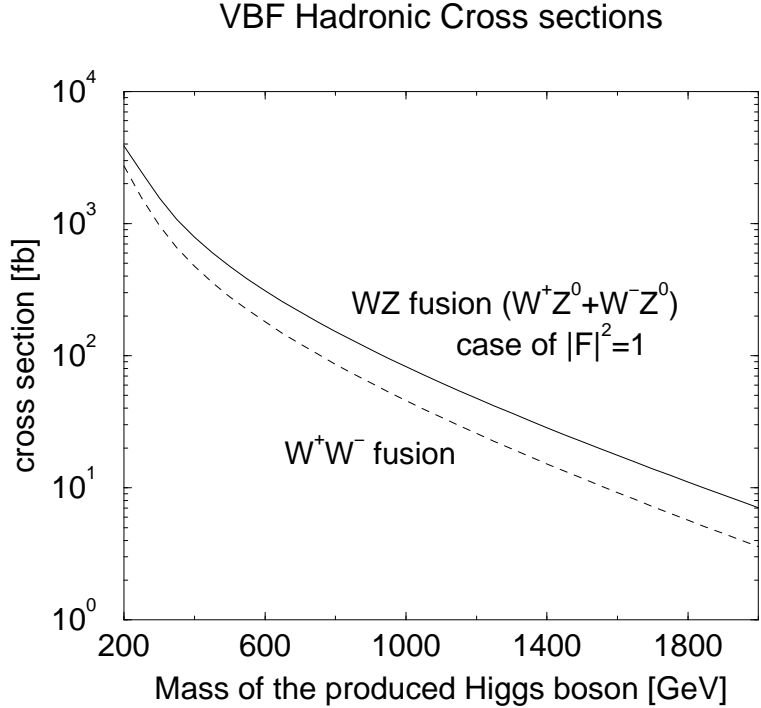}
\caption{The hadronic cross section of 
the $W^\pm Z^0$ fusion process and 
the $W^+W^-$ fusion process as a function of 
the mass of the charged and neutral Higgs bosons, respectively. 
For the $W^\pm Z^0$ fusion, the form factor $F$ is set to be unity.
The SM prediction is shown for the $W^+W^-$ fusion.
}
\label{Fig:VVfusion}
\end{figure}

\subsection{Event selection}
\subsubsection{Higgs production via VBF}
\noindent\underline{Forward jet tagging}

A characteristic of VBF is that the colliding quarks are scattered with 
relatively high $P_T$. These scattered partons can easily be
identified as jets in forward region of the detector.
Since the two forward jets $j_1$ and $j_2$ lie in 
opposite hemispheres with large pseudo-rapidity separation
between them, we can impose the cut
\begin{eqnarray}
\Delta\eta_{j_1 j_2} \equiv |\eta_{j_1}-\eta_{j_2}| > 2-4, \label{eq:eta}
\end{eqnarray}
where the cut value is  adjusted depending on the Higgs mass and decay modes
so as to make it most effectively.
With this cut, QCD backgrounds are typically suppressed
by a few orders of magnitude, while the signal rate is suppressed
at most by a factor of $2-3$. We also impose the constraints for the invariant
mass of two forward jets, whose physical meaning is similar to that in
the constraint in Eq.~(\ref{eq:eta}); 
\begin{eqnarray}
M_{j_1 j_2} > 500-1000~{\rm GeV}.
\end{eqnarray}

\vspace{5mm}
\noindent\underline{Central jet veto}

Since VBF is a pure electroweak process without color flow in the central region,
it allows us to observe small hadronic activity except for jets which are
the tagging jets and
the  decay products
from produced Higgs boson. We then impose the veto on the additional production
of jets
in the central region (between two tagging jets).

\subsubsection{Higgs decay part}

Here we consider the kinematic cuts for
the $H^\pm \to t\overline{b}(\overline{t}b)$ and
the $H^\pm \to W^\pm Z^0$ processes,
separately.
We impose the basic cuts for the invariant mass, and those for  
the number of jets, $b$-jets and leptons which should be the
appropriate number for each final state.  
In order to improve the significance $S/\sqrt{B}$,  
we impose further several effective cuts which we explain below. 

\vspace{5mm}
\noindent\underline{The $H^\pm\ \to t\overline{b}(\overline{t}b)$ mode}

The final state $2b+\ell+\met$ with two tagging jets is studied here. 
The main background is the top pair production whose cross section
amounts to about $490$ pb.

\begin{figure}[t]
\includegraphics[width=12cm]{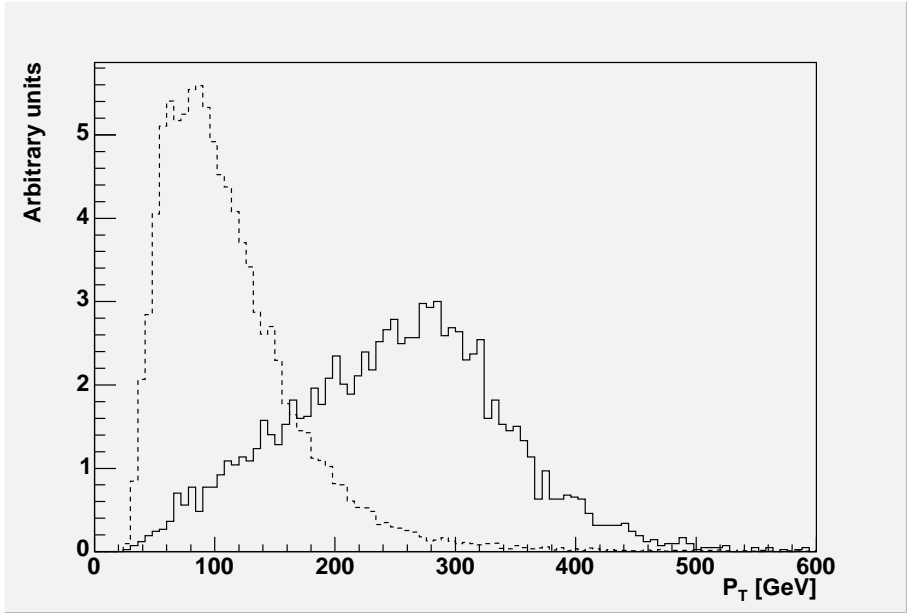}
\caption{The $P_T$ distribution for the $b$-jets with the highest $P_T$
from the signal ($pp \to W^\pm Z^0 X \to H^\pm X \to tbX$ at $m_H^\pm=700$ GeV) and background
(top pair production). Solid (dashed) histogram represents 
the signal (top pair production).}
\label{Fig:ptb}
\end{figure}

For the $H^- \to \bar t b$ mode, 
one $b$-jet ($b$) comes directly from the $H^-$ decay, and the other $b$-jet
($\bar b$) does from the decay of the top quark which is a decay product of $H^-$. 
For $H^-$ with a relatively large mass, the former $b$-jet tends to have
higher $P_T$ than the $b$-jet from the top quark decay. Then,
the cut for the $b$-jet with higher $P_T$ is helpful for reducing the 
top pair production events. The same happens for the $H^+ \to t \bar b$
mode (Fig.~\ref{Fig:ptb}).

\vspace{5mm}
\noindent\underline{The $H^\pm \to W^\pm Z^0$ mode}

For the model with a real and a complex triplet fields~\cite{TRIP,cg,thetaH},
the $H^\pm \rightarrow W^\pm Z^0$ decay mode is dominant
because $H^\pm$ do not couple to fermions.
We consider the sequential decay mode $W^\pm \rightarrow \ell^\pm \nu$ and $Z \rightarrow jj$ 
or $\ell\ell$, that is, the final states are the
$2j+\ell+\met$ and $3\ell+\met$ with two tagging jets.
For the $jj \ell \nu$ mode,
the most serious background comes from the $W+4jets$ events.
Even after the forward jet tagging
($\Delta\eta_{j_1 j_2} > 2.5$ and $M_{j_1 j_2} > 500$ GeV),
the cross section exceeds $130$ pb.
Though the top pair production is also the serious background,
eliminating the events which have two $b$-jets drastically reduces the background
from the top pair production~\cite{zeppenfeld-ww}.

\begin{figure}[t]
\includegraphics[width=10cm]{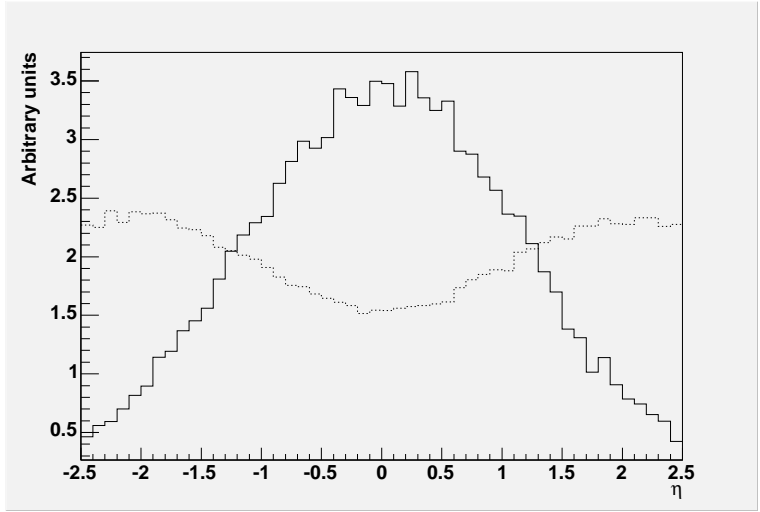}
\caption{The distribution for the pseudo-rapidity $\eta$ of electrons
from the signal ($pp \to W^\pm Z^0 X \to H^\pm X \to W^\pm Z^0 X \to jj \ell \nu X
$ at $m_H^\pm=800$ GeV) 
and backgrounds (top pair production, $W+4j$). Solid (dotted) histogram represents 
the signal ($W+4j$).}
\label{Fig:etael}
\end{figure}

Since the $W+4jets$ events are dominated by single $W$ boson production processes 
$u\overline{d}(\overline{u}d) \rightarrow W^\pm$ at the parton level,
the helicities of the produced $W$ bosons do not have the helicity $0$ component.
Then, the leptons coming from the $W$ bosons are populated in the forward and
backward direction, according to
$|{\cal M}(W^\pm \to \ell^\pm \nu)|^2 \sim  |1 \pm \cos\theta|^2$ where $\theta$ is
the scattering angle in the center-of-mass frame.
On the other hand, $W$ bosons coming from the $H^\pm$ bosons
are not polarized. Then, if we put a cut for the lepton direction,
such as $\eta_\ell<1$ where $\eta_\ell$ is the pseudo-rapidity of the lepton, 
a large amount of backgrounds can be effectively reduced
(Fig.~\ref{Fig:etael}).

\begin{figure}[t]
\includegraphics[width=10cm]{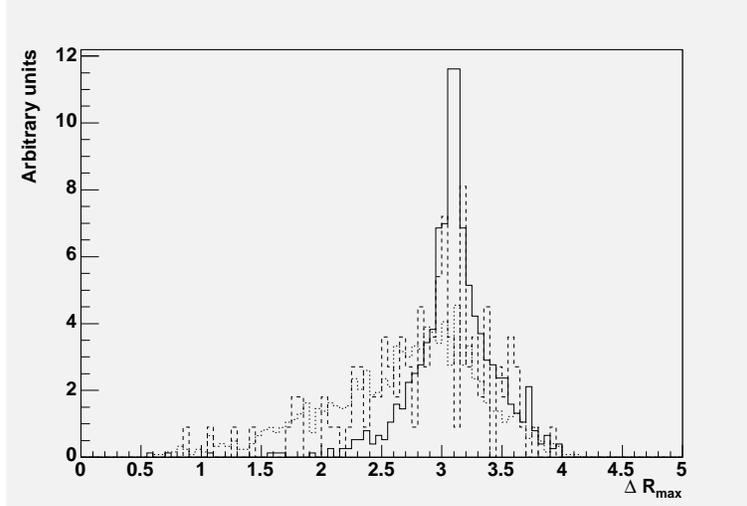}
\caption{The distribution for $\Delta R_{max}$
from the signal ($pp \to W^\pm Z^0 X \to H^\pm X \to W^\pm Z^0 X \to jj \ell \nu X$
at $m_H^\pm=800$ GeV)
and backgrounds
(top pair production, $W+4j$). Solid (dashed, dotted) histogram represents 
the signal (top pair production, $W+4j$).}
\label{Fig:drmax}
\end{figure}

\begin{figure}[h]
\includegraphics[width=10cm]{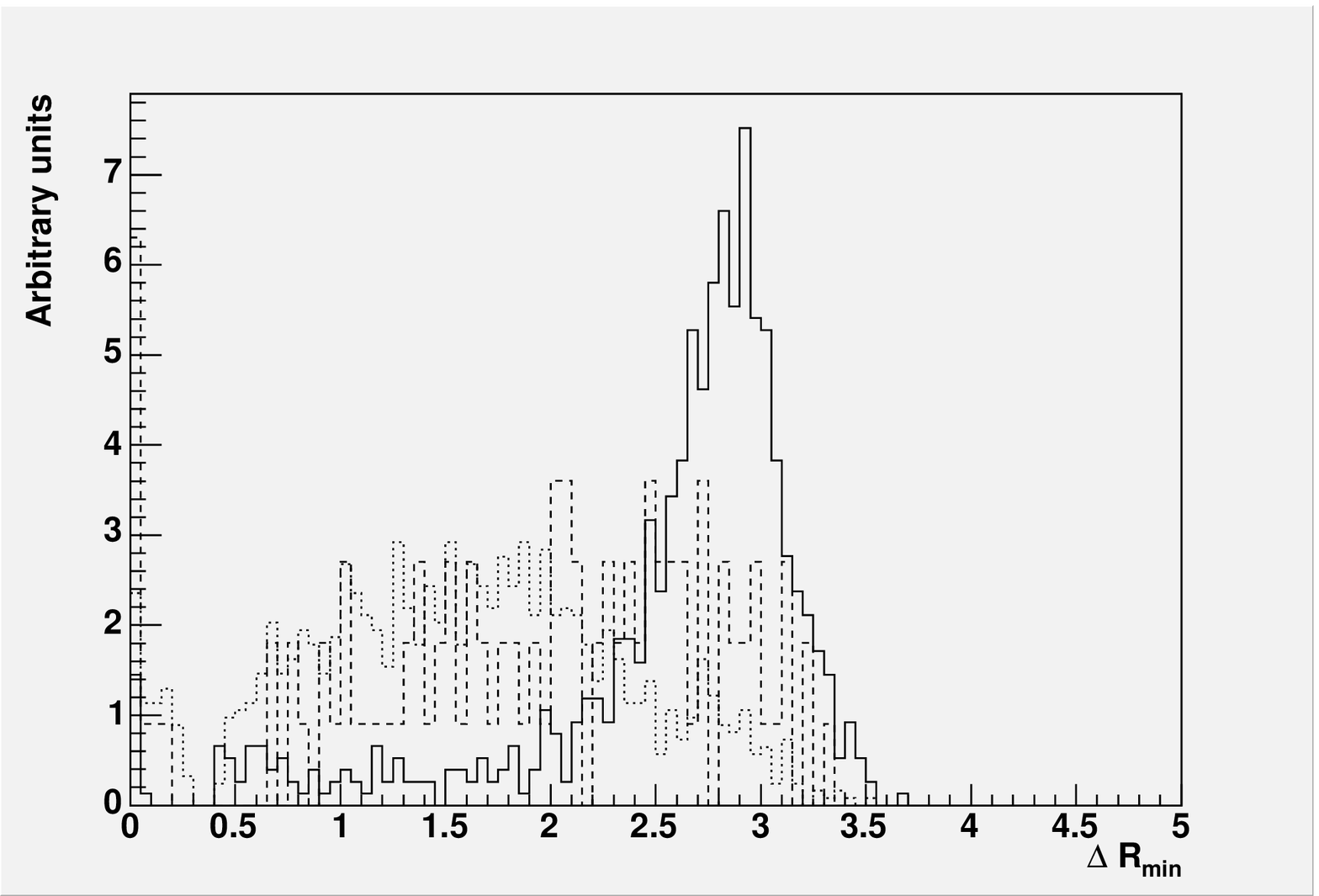}
\caption{The distribution for $\Delta R_{min}$
from the signal ($pp \to W^\pm Z^0 X \to H^\pm X \to W^\pm Z^0 X \to jj \ell \nu X$
at $m_H^\pm=800$ GeV) 
and backgrounds
(top pair production, $W+4j$). Solid (dashed, dotted) histogram represents 
the signal (top pair production, $W+4j$).}
\label{Fig:drmin}
\end{figure}

\begin{figure}[h]
\includegraphics[width=10cm]{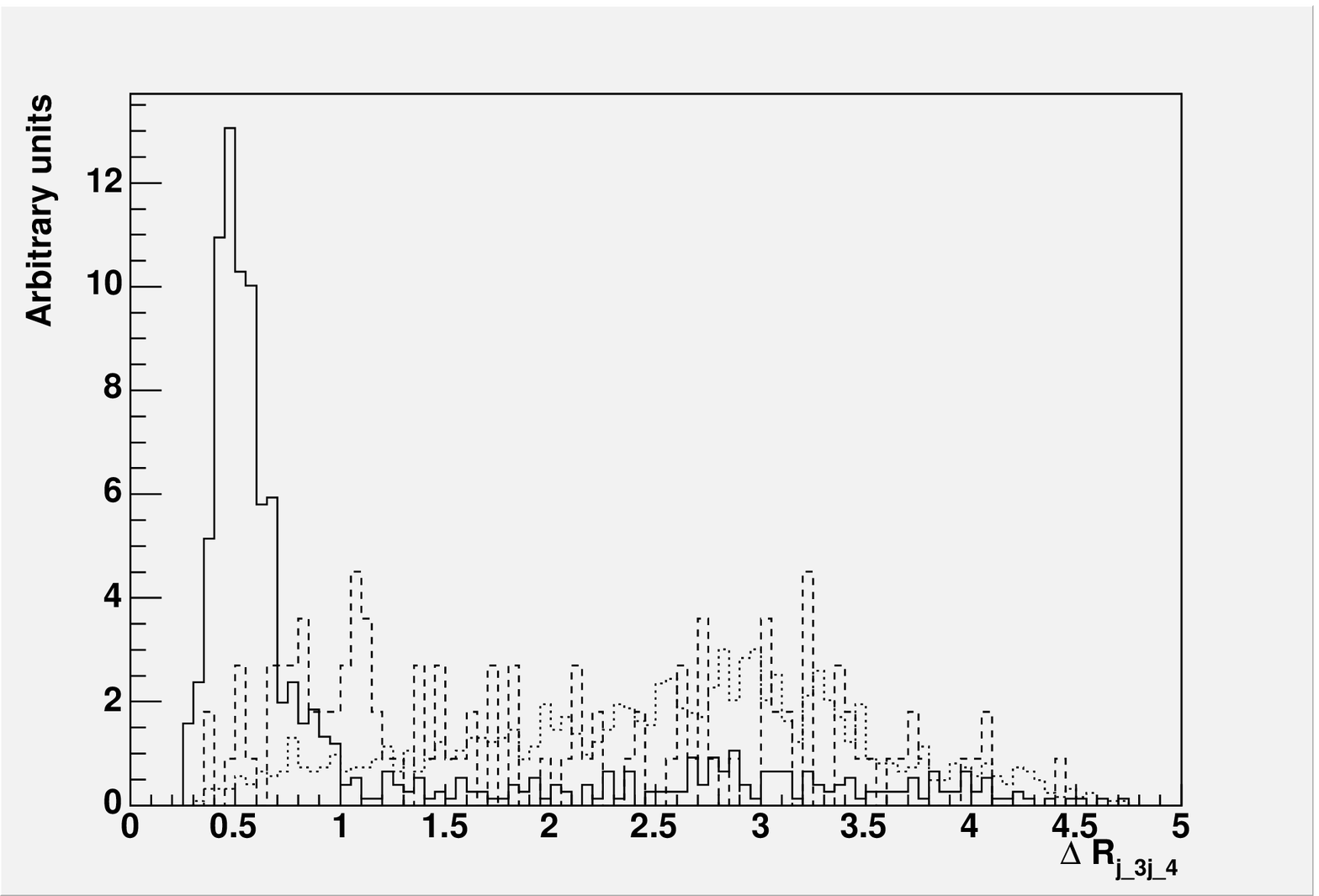}
\caption{The distribution for $\Delta R_{j_3 j_4}$
from the signal ($pp \to W^\pm Z^0 X \to H^\pm X \to W^\pm Z^0 X \to jj \ell \nu X$ 
at $m_H^\pm=800$ GeV) 
and backgrounds
(top pair production, $W+4j$). Solid (dashed, dotted) histogram represents 
the signal (top pair production, $W+4j$).}
\label{Fig:drjj}
\end{figure}

When $H^\pm$ has relatively large masses, the produced $W$ and $Z$ bosons have large momenta
and the trajectories of their decay products follow directions of their parent 
$W$ and $Z$ bosons approximately. Therefore, 
if the jets in the central region are labeled as $j_3$ and $j_4$, and 
if the quantities
$\Delta R_{min} = \min(\Delta R_{\ell j_3}, \Delta R_{\ell j_4})$ and 
$\Delta R_{max} = \max(\Delta R_{\ell j_3}, \Delta R_{\ell j_4})$
are defined,
where
$\Delta R  \equiv \sqrt{|\Delta\phi|^2+|\Delta\eta|^2}$,
the relation
$\Delta R_{min} \simeq \Delta R_{max}$
is obtained for the signal process. 
On the other hand, the background processes
do not necessarily satisfy this relation (Fig.~\ref{Fig:drmax} and Fig.~\ref{Fig:drmin}).
In addition, the jet-jet separation
$\Delta R_{j_3j_4}$
has small value for the signal process(Fig.~\ref{Fig:drjj}), so that
we impose the cut $\Delta R_{j_3 j_4} < 1$.


\subsection{Results}

The event generation for the processes except for $W^\pm Z^0 jj$ production
is performed by PYTHIA 6.2~\cite{pythia62}.\footnote{
We use the neutral Higgs boson production process via VBF as a signal process.
In order to deal with the process of charged Higgs boson production
we compel the produced neutral Higgs bosons to decay into
$t \overline{b}$ ($\overline{t} b$) or $W^\pm Z^0$ in the generation.}
We use the CTEQ5L parametrization~\cite{cteq5l} of the parton distribution
functions.
For the $W^\pm Z^0 jj$ production, we use MadGraph~\cite{madgraph} where
the CTEQ6L~\cite{cteq6l} is adopted.
In our study, we assume the branching ratio of the charged Higgs
decay into each mode ($H^\pm \to t \overline{b}$ ($\overline{t} b$) 
and $H^\pm \to W^\pm Z^0$) 
is $100$\% in order to give results
in a model-independent way. When a particular model is considered,
we have to multiply the branching ratios properly.
For the detector performance we applied acceptance cuts and
included smearing effects according to the ATLAS detector\cite{atlfast}.
For $W+4j$ events, the events with $M_{j_1 j_2} > 500$ GeV and 
$\Delta\eta_{j_1 j_2} > 2.5$ are used in our simulation study.

\subsubsection{$pp \to W^\pm Z^0 X \to H^\pm X \to tbX$}

\noindent\underline{$H^\pm \rightarrow t \overline{b}$ ($\overline{t} b$) 
$\rightarrow b b \ell \nu$}

The results of
the efficiency for selection cuts are
listed in Table~\ref{Htb200} and \ref{Htb700} 
for $m_{H^\pm}=200$ and $700$ GeV, respectively.
The cuts applied in this analysis are:

\begin{itemize}
\item Forward jet tagging: \\
At least one jet in both forward ($\eta \geq 0$)
and backward ($\eta<0$) regions with $P_T>25$ GeV. \\
By defining a jet with the highest $P_T$ in the forward (backward) region
as $j_1$ ($j_2$), \\
$\Delta\eta_{j_1 j_2} > 3.8$, $M_{j_1 j_2} > 550$ ${\rm GeV}$ for $m_{H^\pm}=200$ GeV, \\
$\Delta\eta_{j_1 j_2} > 3.5$, $M_{j_1 j_2} > 800$ ${\rm GeV}$ for $m_{H^\pm}=700$ GeV.

\item Lepton cuts: \\
One lepton with $P_T>30$ GeV.

\item $b$-jet cuts: \\
Two $b$-jets with $P_T>25$ GeV. \\
($P_T$ of one $b$-jet with higher $P_T$) $> 250$ GeV for $m_{H^\pm}=700$ GeV.

\item Central jet veto: \\
No jet with $P_T>20$ GeV and $\Delta R > 0.7$ from tag jets.

\item invariant mass cuts: \\
$M_{bbl\nu} < 220$ GeV for $m_{H^\pm}=200$ GeV.
\end{itemize}

\begin{table}[t]
\begin{center}
\begin{tabular}{l||rr||rr}
& \multicolumn{2}{c}{signal} & \multicolumn{2}{c}{background} \\
&       & & $t \overline{t}$  \\ \hline\hline
Before cuts & 100\%  & (170 fb) & 100\% & (490 pb) \\
+Forward jet tagging & 29\% & (49 fb) & 4.1\% & (20 pb) \\
+Lepton cuts & 4.1\% & (7.0 fb) & 0.81\% & (4.0 pb) \\
+$b$-jet cuts & 1.0\% & (1.7 fb) & 0.42\% & (2.1 pb)  \\
+Central jet veto & 0.66\% & (1.1 fb) & 0.060\% & (0.29 pb) \\
+invariant mass cuts & 0.32\% & (0.54 fb) & 0.0036\% & (18 fb)  
\\ \hline\hline
\end{tabular}
\caption[$m_{H^\pm}=200$ GeV]
{Efficiencies and cross sections for the signal 
($pp \to W^\pm Z^0 X \to H^\pm X \to tbX \to bb \ell \nu X$ at $m_{H^\pm}=200$ GeV)
and the background (top pair production). The cross sections are
shown in parenthesis.
For the signal,
we show the cross sections which give $S/\sqrt{B} \simeq 3$ for $|F|^2=1$.}
\label{Htb200}
\end{center}
\end{table}

\begin{table}[h]
\begin{center}
\begin{tabular}{l||rr||rr}
& \multicolumn{2}{c}{signal} & \multicolumn{2}{c}{background} \\
&       &  & $t \overline{t}$  \\ \hline\hline
Before cuts & 100\%  & (23 fb) & 100\% & (490 pb) \\
+Forward jet tagging & 32\% & (7.4 fb) & 2.1\% & (10.0 pb) \\
+Lepton cuts & 5.5\% & (1.3 fb) & 0.42\% & (2.1 pb) \\
+$b$-jet cuts & 2.1\% & (0.48 fb) & 0.0065\% & (0.032 pb) \\
+Central jet veto & 1.1\% & (0.25 fb) & 0.00079\% & (3.9 fb) \\
\hline\hline
\end{tabular}
\caption[$m_{H^\pm}=700$ GeV]
{Efficiencies and cross sections for the signal ($pp \to W^\pm Z^0 X \to H^\pm X \to tbX
\to bb \ell \nu X$ 
at $m_{H^\pm}=700$ GeV)
and the background (top pair production) for $H^{\pm}\rightarrow 
t \overline{b} (\overline{t} b)$ decay mode. The cross sections are
shown in parenthesis.
For the signal,
we show the cross sections which give $S/\sqrt{B} \simeq 3$ for $|F|^2=1$.}
\label{Htb700}
\end{center}
\end{table}

After these cuts are imposed,
the signal events decrease to 0.32\% (1.1\%) for $m_{H^\pm}=200$ GeV ($700$ GeV),
while a large reduction by about $2$ orders ($3$ orders) of magnitude is observed 
in the background processes.
Since the cross section for the top pair production is $490$ pb,
the signal cross sections of
$170$ fb and $23$ fb are required to
satisfy the statistical significance $S/\sqrt{B} > 3$ for $m_{H^\pm}=200$ and $700$ GeV, 
respectively, assuming the integrated luminosity as ${\cal L}=600$ ${\rm fb}^{-1}$
and the lepton detection efficiency as $90\%$.

\subsubsection{$pp \to W^\pm Z^0 X \to H^\pm X \to W^\pm Z^0 X$}
\underline{$H^\pm \rightarrow W^\pm Z^0 \rightarrow jj \ell \nu$}

Efficiencies for the selection cuts are listed in Table~\ref{HWZ200}, \ref{HWZ500} 
and \ref{HWZ800} 
for $m_{H^\pm}=200$, $500$ and $800$ GeV, respectively.
The cuts applied in this analysis are:

\begin{itemize}
\item Forward jet tagging: \\
At least one jet in both forward ($\eta \geq 0$)
and backward ($\eta<0$) regions with $P_T>40$ GeV.\\
By defining a jet with the highest $P_T$ in the forward (backward) region
as $j_1$ ($j_2$), \\
$\Delta\eta_{j_1 j_2} > 2.5$, $M_{j_1 j_2} > 500$ ${\rm GeV}$.

\item $b$-jet cuts: \\
No $b$-jet with $P_T>25$ GeV. 

\item Lepton cuts: \\
One lepton with $P_T>30$ GeV and no other lepton with $P_T>20$ GeV,\\
$(M_{\ell\nu})_T < 100$ GeV,\\
$\eta_\ell$ $< 1.0$.

\item Central jet veto: \\
Two jets with $P_T>20$ GeV in the central region ($\eta<2.0$).

\item Central jet cuts: \\
($P_T$ of one central jet with higher $P_T$) $> 50$ GeV for $m_{H^\pm}=200$ GeV,\\
($P_T$ of one central jet with higher $P_T$) $> 100$ GeV for $m_{H^\pm}=500$, $800$ GeV,\\
$\Delta R_{j_3 j_4} < 1.0$ and $\Delta R_{max} > 2.5$ for $m_{H^\pm}=500$ GeV,\\
$\Delta R_{j_3 j_4} < 1.0$, $\Delta R_{min} > 2.5$
for $m_{H^\pm}=800$ GeV,\\
$60$ GeV $< M_{j_3 j_4}$ $< 100$ GeV.
\end{itemize}

\begin{table}[h]
\begin{center}
\begin{tabular}{l||rr||rr|rr}
& \multicolumn{2}{c}{signal} & \multicolumn{4}{c}{backgrounds} \\
&        & & \multicolumn{2}{c}{$W+4j$} & \multicolumn{2}{c}{$t \overline{t}$} 
\\ \hline\hline
Before cuts 
& 100\%  & (910 fb) & 100\% & (130 pb) & 100\% & (490 pb)\\
 \hline
+Forward jet tagging & 31\% & (280 fb) & 50\% & (65 pb) & 3.6\% & (18 pb)
\\
 \hline
+$b$-jet cuts & 26\% & (240 fb) & 47\% & (61 pb) & 0.25\% & (1.2 pb) 
\\
 \hline
+Lepton cuts & 1.3\% & (12 fb) & 2.0\% & (2.6 pb) & 0.012\% & (0.059 pb) 
\\
 \hline
+Central jet veto & 0.39\% & (3.5 fb) & 0.32\% & (0.42 pb) & 0.0033\% & (0.016 pb) 
\\
 \hline
+Central jet cuts & 0.11\% & (1.0 fb) & 0.043\% & (56 fb) & 0.00079\% & (3.9 fb) 
\\ 
\hline\hline
\end{tabular}

\begin{tabular}{l||rr||rr|rr|rr}
& \multicolumn{6}{c}{backgrounds} \\
& \multicolumn{2}{c}{$W^+Z^0jj$} & \multicolumn{2}{c}{$W^-Z^0jj$} 
& \multicolumn{2}{c}{$WZ$} \\ \hline\hline
Before cuts 
& 100\% & (350 fb) & 100\% & (210 fb) & 100\% & (26 pb) \\
 \hline
+Forward jet tagging & 43\% & (150 fb) & 36\% & (76 fb)
& 1.1\% & (0.29 pb) \\
 \hline
+$b$-jet cuts & 37\% & (130 fb) & 30\% & (63 fb) 
& 0.99\% & (0.26 pb) \\
 \hline
+Lepton cuts & 1.5\% & (5.3 fb) & 1.2\% & (2.5 fb) 
& 0.015\% & (0.0039 pb) \\
 \hline
+Central jet veto & 0.40\% & (1.4 fb) & 0.30\% & (0.63 fb) 
& 0.0031\% & (0.81 fb) \\
 \hline
+Central jet cuts & 0.081\% & (0.28 fb) & 0.072\% & (0.15 fb) 
& 0.0006\% & (0.16 fb)\\ 
\hline\hline
\end{tabular}
\caption[$m_{H^\pm}=200$ GeV]
{Efficiencies and cross sections for the signal 
($pp \to W^\pm Z^0 X \to H^\pm X \to W^\pm Z^0 X
\to jj \ell\nu X$ at $m_H^\pm=200$ GeV)
and backgrounds ($W+4j$, top pair production, $W^+Z^0jj$, $W^-Z^0jj$, $WZ$ production).
The cross sections are
shown in parenthesis.
For the signal,
we show the cross sections which give $S/\sqrt{B} \simeq 3$ for $|F|^2=1$.}
\label{HWZ200}
\end{center}
\end{table}

\begin{table}[t]
\begin{center}
\begin{tabular}{l||rr||rr|rr}
& \multicolumn{2}{c}{signal} & \multicolumn{4}{c}{backgrounds} \\
&      & & \multicolumn{2}{c}{$W+4j$} & \multicolumn{2}{c}{$t \overline{t}$} 
\\ \hline\hline
Before cuts 
& 100\% & (140 fb) & 100\% & (130 pb) & 100\% & (490 pb) \\
\hline
+Forward jet tagging & 40\% & (56 fb) & 50\% & (65 pb) & 3.6\% & (18 pb) 
\\
\hline
+$b$-jet cuts & 33\% & (46 fb) & 47\% & (61 pb) & 0.25\% & (1.2 pb) 
\\
\hline
+Lepton cuts & 2.0\% & (2.8 fb) & 2.0\% & (2.6 pb) & 0.012\% & (0.059 pb) 
\\
\hline
+Central jet veto & 0.62\% & (0.87 fb) & 0.32\% & (0.42 pb) & 0.0033\% & (0.016 pb) 
\\
\hline
+Central jet cuts & 0.19\% & (0.27 fb) & 0.0027\% & (3.5 fb) & 0.0001\% & (0.49 fb) 
\\
\hline\hline
\end{tabular}

\begin{tabular}{l||rr||rr|rr|rr}
& \multicolumn{6}{c}{backgrounds} \\
& \multicolumn{2}{c}{$W^+Z^0jj$} & \multicolumn{2}{c}{$W^-Z^0jj$} 
& \multicolumn{2}{c}{$WZ$} \\ \hline\hline
Before cuts 
& 100\% & (350 fb) & 100\% & (210 fb) & 100\% & (26 pb) \\
 \hline
+Forward jet tagging & 43\% & (150 fb) & 36\% & (76 fb)
& 1.1\% & (0.29 pb) \\
 \hline
+$b$-jet cuts & 37\% & (130 fb) & 30\% & (63 fb) 
& 0.99\% & (0.26 pb) \\
 \hline
+Lepton cuts & 1.5\% & (5.3 fb) & 1.2\% & (2.5 fb) 
& 0.015\% & (0.0039 pb) \\
 \hline
+Central jet veto & 0.40\% & (1.4 fb) & 0.30\% & (0.63 fb) 
& 0.0031\% & (0.81 fb) \\
 \hline
+Central jet cuts & 0.018\% & (0.063 fb) & 0.022\% & (0.046 fb) 
& 0.0001\% & (0.026 fb)\\ 
\hline\hline
\end{tabular}
\caption[$m_{H^\pm}=500$ GeV]
{Efficiencies and backgrounds for the signal 
($pp \to W^\pm Z^0 X \to H^\pm X \to W^\pm Z^0 X
\to jj \ell\nu X$ at $m_H^\pm=500$ GeV) 
and backgrounds ($W+4j$, top pair production, $W^+Z^0jj$, $W^-Z^0jj$, $WZ$ production). 
The cross sections are shown in parenthesis.
For the signal,
we show the cross sections which give $S/\sqrt{B} \simeq 3$ with $|F|^2=1$.}
\label{HWZ500}
\end{center}
\end{table}

\begin{table}[t]
\begin{center}
\begin{tabular}{l||rr||rr|rr}
& \multicolumn{2}{c}{signal} & \multicolumn{4}{c}{backgrounds} \\
&       & & \multicolumn{2}{c}{$W+4j$} & \multicolumn{2}{c}{$t \overline{t}$} 
 \\ \hline\hline
Before cuts 
& 100\% & (51 fb) & 100\% & (130 pb) & 100\% & (490 pb) \\
 \hline
+Forward jet tagging & 46\% & (23 fb) & 50\% & (65 pb) & 3.6\% & (18 pb) 
\\
 \hline
+$b$-jet cuts & 38\% & (19 fb) & 47\% & (61 pb) & 0.25\% & (1.2 pb) 
\\
 \hline
+Lepton cuts & 2.5\% & (1.3 fb) & 2.0\% & (2.6 pb) & 0.012\% & (0.059 pb) 
\\
 \hline
+Central jet veto & 0.88\% & (0.45 fb) & 0.32\% & (0.42 pb) & 0.0033\% & (0.016 pb) 
\\
 \hline
+Central jet cuts & 0.42\% & (0.21 fb) & 0.0016\% & (2.1 fb) & 0.0001\% & (0.49 fb) 
\\
\hline\hline
\end{tabular}

\begin{tabular}{l||rr||rr|rr|rr}
& \multicolumn{6}{c}{backgrounds} \\
& \multicolumn{2}{c}{$W^+Z^0jj$} & \multicolumn{2}{c}{$W^-Z^0jj$} 
& \multicolumn{2}{c}{$WZ$} \\ \hline\hline
Before cuts 
& 100\% & (350 fb) & 100\% & (210 fb) & 100\% & (26 pb) \\
 \hline
+Forward jet tagging & 43\% & (150 fb) & 36\% & (76 fb)
& 1.1\% & (0.29 pb) \\
 \hline
+$b$-jet cuts & 37\% & (130 fb) & 30\% & (63 fb) 
& 0.99\% & (0.26 pb) \\
 \hline
+Lepton cuts & 1.5\% & (5.3 fb) & 1.2\% & (2.5 fb) 
& 0.015\% & (0.0039 pb) \\
 \hline
+Central jet veto & 0.40\% & (1.4 fb) & 0.30\% & (0.63 fb) 
& 0.0031\% & (0.81 fb) \\
 \hline
+Central jet cuts & 0.014\% & (0.049 fb) & 0.018\% & (0.038 fb) 
& 0.0001\% & (0.026 fb)\\ 
\hline\hline
\end{tabular}
\caption[$m_{H^\pm}=800$ GeV]
{Efficiencies and backgrounds for the signal 
($pp \to W^\pm Z^0 X \to H^\pm X \to W^\pm Z^0 X
\to jj \ell\nu X$ at $m_H^\pm=800$ GeV) 
and backgrounds ($W+4j$, top pair production, $W^+Z^0jj$, $W^-Z^0jj$, $WZ$ production)
for . The cross sections are
shown in parenthesis.
For the signal,
we show the cross sections which give $S/\sqrt{B} \simeq 3$ for $|F|^2=1$.}
\label{HWZ800}
\end{center}
\end{table}

The event selection also works effectively in order to reduce 
the $WZ$ production whose cross section
is 26 pb. According to our simulation study, the number of remaining events
for the $WZ$ and $WZjj$ production processes is 
less than the one for $W+4jet$ production by at least 2 orders.
  
Since the cross section for the $W+4jet$ production (the top pair production) is about $130$ pb
($490$ pb), the signal cross sections of 
$910$ fb, $140$ fb and $51$ fb are required
to satisfy $S/\sqrt{B} > 3$ for $m_{H^\pm}=200$, 
$500$ and $800$ GeV, 
respectively, under ${\cal L}=600$ ${\rm fb}^{-1}$ and the $90\%$ lepton detection efficiency.

\vspace{5mm}
\underline{$H^\pm \rightarrow W^\pm Z^0 \rightarrow \ell\ell\ell \nu$}

Efficiencies for the selection cuts are listed in Table~\ref{HWZlvll200}, \ref{HWZlvll500} 
and \ref{HWZlvll800} 
for $m_{H^\pm}=200$, $500$ and $800$ GeV, respectively.
The cuts applied in this analysis are:

\begin{itemize}
\item Forward jet tagging: \\
At least one jet in both forward ($\eta \geq 0$)
and backward ($\eta<0$) regions with $P_T>40$ GeV.\\
By defining a jet with the highest $P_T$ in the forward (backward) region
as $j_1$ ($j_2$), \\
$\Delta\eta_{j_1 j_2} > 2.5$, $M_{j_1 j_2} > 500$ ${\rm GeV}$.

\item Lepton cuts: \\
Three leptons with $P_T>30$ GeV and no other lepton with $P_T>20$ GeV.\\
By defining two leptons whose combination minimizes the quantity
$M_{\ell_i \ell_j}-m_Z$ ($i,j = 1-3$, $m_Z$ is the mass of the $Z$ boson.)
as $\ell_{Z_1}$ and $\ell_{Z_2}$,\\
$\Delta R_{\ell_{Z_1} \ell_{Z_2}} < 1.0$ for $m_{H^\pm}=800$ GeV,\\
$80$ GeV $< M_{\ell_{Z_1} \ell_{Z_2}} < 100$ GeV,\\
$170$ GeV $< M_{\ell \ell \ell \nu} < 240$ GeV for $m_{H^\pm}=200$ GeV,\\
$450$ GeV $< M_{\ell \ell \ell \nu} < 600$ GeV for $m_{H^\pm}=500$ GeV.\\
\end{itemize}

\begin{table}[t]
\begin{center}
\begin{tabular}{l||rr||rr|rr|rr}
& \multicolumn{2}{c}{signal} & \multicolumn{6}{c}{backgrounds} \\
&        & & \multicolumn{2}{c}{$W^+Z^0jj$} & \multicolumn{2}{c}{$W^-Z^0jj$} 
& \multicolumn{2}{c}{$WZ$} \\ \hline\hline
Before cuts 
& 100\%  & (210 fb) & 100\% & (350 fb) & 100\% & (210 fb) & 100\% & (26 pb) \\
 \hline
+Forward jet tagging & 31\% & (65 fb) & 43\% & (150 fb) & 36\% & (76 fb)
& 1.1\% & (0.29 pb) \\
 \hline
+Lepton cuts & 0.017\% & (0.036 fb) & 0.0040\% & (0.014 fb) & 0.0040\% & (0.0084 fb) 
&  & ($<0.0026$ fb) \\
\hline\hline
\end{tabular}
\caption[$m_{H^\pm}=200$ GeV]
{Efficiencies and cross sections for the signal 
($pp \to W^\pm Z^0 X \to H^\pm X \to W^\pm Z^0 X
\to \ell \ell \ell \nu X$ at $m_H^\pm=200$ GeV)
and backgrounds ($W^+Z^0jj$, $W^-Z^0jj$, $WZ$ production).
The cross sections are
shown in parenthesis.
For the signal,
we show the cross sections which give $S/\sqrt{S+B} \simeq 3$ for $|F|^2=1$.}
\label{HWZlvll200}
\end{center}
\end{table}

\begin{table}[t]
\begin{center}
\begin{tabular}{l||rr||rr|rr|rr}
& \multicolumn{2}{c}{signal} & \multicolumn{6}{c}{backgrounds} \\
&        & & \multicolumn{2}{c}{$W^+Z^0jj$} & \multicolumn{2}{c}{$W^-Z^0jj$} 
& \multicolumn{2}{c}{$WZ$} \\ \hline\hline
Before cuts 
& 100\%  & (110 fb) & 100\% & (350 fb) & 100\% & (210 fb) & 100\% & (26 pb) \\
 \hline
+Forward jet tagging & 40\% & (44 fb) & 43\% & (150 fb) & 36\% & (76 fb)
& 1.1\% & (0.29 pb) \\
 \hline
+Lepton cuts & 0.030\% & (0.033 fb) & 0.0040\% & (0.014 fb) &  & ($<0.0042$ fb) 
&  & ($<0.0026$ fb) \\
\hline\hline
\end{tabular}
\caption[$m_{H^\pm}=500$ GeV]
{Efficiencies and cross sections for the signal 
($pp \to W^\pm Z^0 X \to H^\pm X \to W^\pm Z^0 X
\to \ell \ell \ell \nu X$ at $m_H^\pm=500$ GeV)
and backgrounds ($W^+Z^0jj$, $W^-Z^0jj$, $WZ$ production).
The cross sections are
shown in parenthesis.
For the signal,
we show the cross sections which give $S/\sqrt{S+B} \simeq 3$ for $|F|^2=1$.}
\label{HWZlvll500}
\end{center}
\end{table}

\begin{table}[t]
\begin{center}
\begin{tabular}{l||rr||rr|rr|rr}
& \multicolumn{2}{c}{signal} & \multicolumn{6}{c}{backgrounds} \\
&        & & \multicolumn{2}{c}{$W^+Z^0jj$} & \multicolumn{2}{c}{$W^-Z^0jj$} 
& \multicolumn{2}{c}{$WZ$} \\ \hline\hline
Before cuts 
& 100\%  & (67 fb) & 100\% & (350 fb) & 100\% & (210 fb) & 100\% & (26 pb) \\
 \hline
+Forward jet tagging & 46\% & (31 fb) & 43\% & (150 fb) & 36\% & (76 fb)
& 1.1\% & (0.29 pb) \\
 \hline
+Lepton cuts & 0.090\% & (0.060 fb) & 0.026\% & (0.091 fb) & 0.010\% & (0.021 fb) 
& & ($<0.0026$ fb) \\
\hline\hline
\end{tabular}
\caption[$m_{H^\pm}=800$ GeV]
{Efficiencies and cross sections for the signal 
($pp \to W^\pm Z^0 X \to H^\pm X \to W^\pm Z^0 X
\to \ell \ell \ell \nu X$ at $m_H^\pm=800$ GeV)
and backgrounds ($W^+Z^0jj$, $W^-Z^0jj$, $WZ$ production).
The cross sections are
shown in parenthesis.
For the signal,
we show the cross sections which give $S/\sqrt{S+B} \simeq 3$ for $|F|^2=1$.}
\label{HWZlvll800}
\end{center}
\end{table}
  
Since the cross section for the $W^+Z^0jj$ production (the $W^-Z^0jj$ production) 
is about $350$ fb ($210$ fb), the signal cross sections of 
$210$ fb, $110$ fb and $67$ fb are required
to satisfy $S/\sqrt{S+B} > 3$ for $m_{H^\pm}=200$, 
$500$ and $800$ GeV, 
respectively, under ${\cal L}=600$ ${\rm fb}^{-1}$ and the $90\%$ lepton detection efficiency.

\section{Discussions}

\label{Sec:WZfusion}



\subsection{$pp \to W^\pm Z^0 X \to H^\pm X \to tbX$}
In the 2HDM and the MSSM, although there are potentially many decay
modes, the main mode would be the decay into 
a $t b$ pair as long as it is kinematically allowed.
The main decay mode of the Littlest Higgs model
would also be the decay into a $tb$ pair~\cite{LH-ph}.
In these cases, the signal can be   
a $b b \ell \nu$ ($\ell=e$ and $\mu$) event.
As shown in Sec.~III, the required cross section
to satisfy $S/\sqrt{B} \gsim 3$ after background reduction
is $\sigma_S^{} \gsim 170$ fb ($23$ fb) for $m_{H^\pm}=200$ GeV
($700$ GeV), which corresponds to 
$|F|^2 \gsim 0.044$  ($|F|^2 \gsim 0.11$),
where $\sigma_S$ is the signal cross section before kinematical cuts.   
Hence, the typical values of $|F|^2$ in the 2HDM,
the MSSM and the Littlest Higgs model
are all smaller than the required one for $S/\sqrt{B} \gsim
3$\footnote{When the variation of the models such as the Littlest Higgs model
with only gauging the SM $U(1)_Y$ subgroup~\cite{csaki} are considered,
the $F$ values can be larger. Then, the observation may be possible.}.
Therefore, we must conclude that
testing these models through this process is challenging. 

\subsection{$pp \to W^\pm Z^0 X \to H^\pm X \to W^\pm Z^0 X$}

In models with triplets that
do not couple to fermions, it would mainly decay into a $WZ$ pair. 
The model with a real and a complex triplets 
can correspond to this case. 
The signal event can be $jj \ell \nu$ and $\ell \ell \ell \nu$.

For the $jj \ell \nu$ event,
the required values of the production cross section
to satisfy the statistical significance $S/\sqrt{B} \gsim 3$ after background reduction
is $\sigma_S^{} \gsim 910$, $140$ and $51$ fb for $m_{H^\pm}=200$, $500$
and $800$ GeV, respectively.
The corresponding values of $|F|^2$ are  
$|F|^2 \gsim 0.23$, $0.30$ and $0.34$.
For the $\ell\ell\ell \nu$ event,
the required values of the production cross section
to satisfy the statistical significance $S/\sqrt{S+B} \gsim 3$ after background reduction
is $\sigma_S^{} \gsim 210$, $110$ and $67$ fb for $m_{H^\pm}=200$, $500$
and $800$ GeV, respectively.
The corresponding values of $|F|^2$ are  
$|F|^2 \gsim 0.054$, $0.23$ and $0.45$. The results are similar to
those in Ref.~\cite{higgsless}.

Therefore, this model can be tested via the process 
$pp \to W^\pm Z^0 X \to H^\pm X \to W^\pm Z^0 X$.

\begin{table}[t]
\begin{center}
\begin{tabular}{|l||l|l|}
\hline
$m_H^\pm$ (GeV) & ~200 & ~700 \\ \hline\hline 
 $|F|^2$ for $bb\ell\nu$ & 0.044 & 0.11~ \\
 \hline
\end{tabular}
\caption
{Required $|F|^2$ values for $S/\sqrt{B} \sim 3$ 
for the $pp \to W^\pm Z^0 X \to H^\pm X \to tbX
\to b b \ell \nu$
mode at ${\cal L}=600$ $fb^{-1}$.}
\label{summary1}
\end{center}
\end{table}

\begin{table}[tb]
\begin{center}
\begin{tabular}{|l||l|l|l|}
\hline
$m_H^\pm$ (GeV) & ~200 & ~500 & ~800 \\ \hline\hline 
 $|F|^2$ for $jj\ell\nu$ & 0.23 & 0.30 & 0.34 \\ \hline
 $|F|^2$ for $\ell\ell\ell\nu$ & 0.054 & 0.23~ & 0.45~ \\
 \hline
\end{tabular}
\caption
{Required $|F|^2$ values for $S/\sqrt{B} \sim 3$ 
for the $pp \to W^\pm Z^0 X \to H^\pm X \to W^\pm Z^0 X \to
jj\ell\nu X$ and $\ell\ell\ell\nu X$ modes
at ${\cal L}=600$ $fb^{-1}$.}
\label{summary2}
\end{center}
\end{table}
\section{Conclusions}
\label{Sec:conclusions}

 The $H^\pm W^\mp Z^0$ vertex strongly depends
 on the structure of the Higgs sector in various new physics scenarios,
 so that its measurement can be useful to distinguish the models. 
 In this paper, the possibility of measuring this 
 vertex has been studied 
 by using the single $H^\pm$ production via $WZ$ fusion
 at the LHC.
 A signal and background simulation under the expected detector
 performance at the LHC is performed.
Required values of $|F|^2$ for $S/\sqrt{B} \gsim 3$ are obtained 
for the $pp \to W^\pm Z^0 X \to H^\pm X \to tbX \to b b \ell \nu X$ mode and 
the $pp \to W^\pm Z^0 X \to H^\pm X \to W^\pm Z^0 X \to
jj\ell\nu X$ and $\ell\ell\ell\nu X$ modes 
:see Tables~\ref{summary1} and \ref{summary2}, respectively.

The process of $pp \to W^\pm Z^0 X \to H^\pm X \to tbX \to bb \ell \nu X$
can be used to test $H^\pm W^\mp Z^0$ vertex
in multi-Higgs doublet models, the MSSM
as well as the Littlest Higgs model. 
However, the required magnitude of the form factor $F$
for the observation via $WZ$ fusion at the LHC is turned out to be
above the predicted maximal values of $|F|^2$ in these models.
The maximal value of $|F|^2$ in the 2HDM is less by 2 orders 
than the required magnitude of $|F|^2$,
and in the Littlest Higgs model
the predicted $|F|^2$ values are less than the required $|F|^2$ value 
by 1 order. If we go to the SLHC, the $10$ times larger luminosity may be expected.
However, it is not easy to observe because the $b$ tagging efficiency should be
much worse than that of the LHC.
Finally, although we have concentrated on $WZ$ fusion in this paper,
the $H^\pm W^\mp Z^0$ coupling may also be studied by using
$gb \to H^\pm t \to W^\pm Z^0 t$ for models which include the sizable coupling
of $H^\pm t b$ like in the 2HDM if the background can be greatly reduced.

On the other hand, it turns out that the process 
of $pp \to W^\pm Z^0 X \to H^\pm X \to W^\pm Z^0 X \to \ell \ell \ell \nu X$
and $jj \ell \nu X$ can be useful to test the model
with a real and a complex triplet fields.

\vspace{1cm}
\noindent
{\large \it Acknowledgments}

The authors would like to thank 
Shoji Asai, Tomio Kobayashi and Mihoko Nojiri 
for useful discussions.  
A part of this work started in the discussion 
during the workshop ``Physics in LHC era'' at YITP, 13-15 
December 2004 (YITP-W-04-20). 
S.K. was supported, in part, by Grants-in-Aid of the Ministry 
of Education, Culture, Sports, Science and Technology, Government of 
Japan, Grant Nos. 17043008 and 18034004.


\end{document}